# Observations on the Halting Problem

Eric C.R. Hehner

Department of Computer Science, University of Toronto
hehner@cs.utoronto.ca

**Abstract:** The Halting Problem is ill-conceived and ill-defined.

## Halting Problem

When Alan Turing laid the foundation for computation in 1936 [5], he wanted to show what computation can do, and what it cannot do. For the latter, he invented a problem that we now call the "Halting Problem". In modern terms, it is as follows.
> In a general-purpose programming language, write a program that reads a text (character string) $p$ representing a program in that same language, and reads another text $i$ representing its input, and outputs *true* if execution of $p$ with input $i$ terminates, and outputs *false* if execution of $p$ with input $i$ does not terminate.

The choice of programming language does not matter; any general-purpose programming language will do. The problem cannot be solved; there is no such program.

Is the problem well defined? It certainly sounds well defined. The input is clear: two texts. The first text is supposed to represent a program; whether it does can presumably be determined the same way a compiler determines whether its input text represents a program. The output is clear: either *true* or *false*. The criterion for outputting *true* is "execution of $p$ with input $i$ terminates", and the criterion for outputting *false* is its negation. What could be clearer?

The problem talks about a "program" to compute halting, but for convenience, without loss of generality, I will talk about a Pascal function to compute halting. The problem talks about whether execution of a "program" terminates, but for convenience, without loss of generality, I will talk about whether execution of a Pascal procedure terminates. The argument for incomputability begins with the assumption, made for the sake of showing a contradiction, that the halting function can be programmed, and has been programmed; let's call it *halts*. Then we can write a procedure like this:

```
procedure diag (s: string);
begin
        if halts (s, s) then diag (s)
end
```

Does execution of *diag* (*'diag'*) terminate? If it does, then *halts* (*'diag'*, *'diag'*) should return *true*, and so we see from the body of procedure *diag* that its execution does not terminate. If it doesn't terminate, then *halts* (*'diag'*, *'diag'*) should return *false*, and so we see from the body of procedure *diag* that its execution terminates. We have a contradiction (inconsistency), so we conclude that the initial assumption was wrong: the halting function cannot be programmed; it is incomputable.

That argument, and its conclusion, are well accepted. But I have three complaints. My first complaint is small and easily fixed. My middle complaint is more serious, and casts doubt on the conclusion. My final complaint casts doubt on the whole problem.

## Domain Problem

Execution of *diag* (*'diag'*) includes function call *halts* (*'diag'*, *'diag'*) . Function *halts* requires its first argument *'diag'* to be a text that represents a legal (syntactically correct and type correct) procedure. Suppose *diag* is a legal procedure; then the arguments of *halts* are in their domains, and so the call *halts* ('*diag*', '*diag*') is legal, and therefore *diag* is a legal procedure, as supposed. Suppose *diag* is not a legal procedure; then the first argument of *halts* is not in its domain, and so the call *halts* ('*diag*', '*diag*') is illegal, and therefore *diag* is not a legal procedure, as supposed. We are unable to say whether *diag* is a legal procedure.

The solution to this domain problem is simple and surprising. Change the Halting Problem as follows.
> In a general-purpose programming language, write a program that reads two texts (character strings) *p* and *i* . If *p* represents a program in that same language whose execution with input *i* terminates, then output *true* . If *p* does not represent a program in that same language, or represents a program in that same language whose execution with input *i* does not terminate, then output *false* .

Now the call *halts* ('*diag*', '*diag*') is legal no matter whether *diag* is a legal procedure; therefore, ironically, *diag* is a legal procedure.

## Specification Problem

The next problem is more serious. When we reason about the execution of procedure *diag* , how do we know what the function call *halts* ('*diag*', '*diag*') will return? In general, there are two ways. The way preferred by the semantics and verification communities is to examine the program for function *halts* . The incomputability argument begins with the assumption that we have programmed function *halts* , so that procedure *diag* can call it and it can be executed. But we do not actually have the program for function *halts* for the purpose of examining it to determine what *halts* ('*diag*', '*diag*') returns. Under the present circumstance, this way doesn't work.

The other way is to examine the specification of function *halts* as stated in the Halting Problem. As a programmer, I consider that the meaning of a call is always given by its specification, even if the program is available. Under the present circumstance, this is the only possible option, and this is how the incomputability proof proceeds. But we don't need to assume that halting is computable, or that it has been programmed, to use the specification in our reasoning. Without assuming computability, we ask what the result of *halts* ('*diag*', '*diag*') should be. This is a question about the specification of *halts* . If it should be *true* , then the semantics of *diag* ('*diag*') is nontermination, so *halts* ('*diag*', '*diag*') should be *false* . If it should be *false* , then the semantics of *diag* ('*diag*') is termination, so it should be *true* . This is inconsistent. Therefore *halts* cannot be programmed according to its inconsistent specification.

It is difficult to see how the *halts* specification could be inconsistent, but the *diag* example shows us that it is. We arrive at the conclusion that *halts* cannot be programmed, but the reason is not incomputability of a well-defined mathematical function. The reason is that the specification of the *halts* function is inconsistent. The question of computability of a well-defined mathematical function has not been addressed.

We could also define

**procedure** *what* (*s*: **string**);
**begin**
    **if not** *halts* (*s*, *s*) **then** *what* (*s*)
**end**

and then ask what the result of *halts* ('*what*', '*what*') should be. If it should be *true* , then the semantics of *what* ('*what*') is termination, so *halts* ('*what*', '*what*') should be *true* , as assumed. If it should be *false* , then the semantics of *what* ('*what*') is nontermination, so it should be *false* , as assumed. Both answers are acceptable. Even though the informal specification of *halts* seemed clear, it is overdetermined when applied to *diag* , and underdetermined when applied to *what* .

## Meaning Problem

Suppose you buy some software, and it comes with a guarantee: all executions of this software terminate, or one million times your money back. Sounds good. Under what circumstances can you complain that the guarantee has been violated, and demand the promised money? No matter how long the execution has taken, the vendor can say "wait longer". There is never a time when you can say that the computation has taken forever. The guarantee of termination is perfectly safe for the vendor, and worthless for the buyer.

Suppose you are commissioned to write some software, and the client has specified that all executions of the software must terminate. You can safely ignore this specification for the reason stated in the previous paragraph: the client can never complain that the specification has not been met. If the guarantee, or the specification, had stated a time bound within which the termination must occur, then it cannot be ignored because a violation is observed when the time bound is exceeded. But the guarantee and specification said only "termination", without stating a time bound, making them worthless.

The philosopher Karl Popper [2] said that a scientific statement is meaningful if and only if
- it makes sense to say that it is true (it is not self-contradictory), and
- it makes sense to say that it is false (it is not a tautology), and
- if it is false, there is a way to show that it is false (its falsity can be observed).

The fact that nontermination is unobservable makes termination meaningless, according to this philosophy.

In the same vein, according to Shannon's Information Theory [4], a test that cannot fail conveys no information. Similarly, according to Bayesian probability, we cannot confirm something (increase its probability) without a test that can potentially disconfirm it (decrease its probability).

I am hesitant to declare what is meaningful and what is meaningless. But I am prepared to say that if there is no possibility to collect on a guarantee, then the guarantee is worthless. Similarly, if there is no possibility to violate a specification, then the specification is worthless. The halting specification is certainly worthless, possibly meaningless.

In some computing models, termination is observable: it is an event that cannot happen before the end of a computation, and must happen at its end. In these models, nontermination is a worthwhile claim or guarantee or specification. For many applications, such as the control of a nuclear power plant, or a heart pacemaker, it is essential that execution of the software not terminate.

In other models of computation, termination is not a computation event, but the end of or cessation of computation events. In these models, termination is unobservable because you can never be sure that the computation events have ceased. Promising or specifying nontermination, in these models, is as worthless as promising or specifying termination.

I have been discussing the worthlessness of specifying termination based on our inability to observe nontermination when executing a program. It may seem there is another way to approach the problem: analyze, rather than execute, the program. This is particularly relevant for the Halting Problem; function *halts* is given the text of a program for analysis. If you want to prove termination, or prove nontermination, the proof is conducted within a theory, which provides the axioms and proof rules used in the proof. So we need a theory of programming, and four spring to mind:
- Hoare Logic [1], the original 1969 theory of programming. pre- postconditions, invariants, variants.
- Unroll loops and recursions, form a sequence of finite approximations to the behavior, and then take the limit.
- Write recursive equations, then find the least fixed-point solution (the least deterministic solution).

- My own theory of programming (aPToP) [0]. The meaning of a function or procedure call is the function or procedure specification, not the function or procedure body. A loop is syntactic sugar for a recursive call, and the meaning of that call is the loop specification, not the loop body. You don't need to find an invariant; you don't need to form a sequence of approximations; you don't need to calculate a fixed-point.

These theories have different strengths, roughly increasing in the order listed. If ever two theories disagree about what happens when a program is executed, we would arbitrate by executing the program, observe what happens, and throw away the theory that's wrong. The correctness (soundness, validity) of a theory is decided by observation of execution, so proof does not dispense with the need for observation. Fortunately, whenever these theories say something that can be observed about a computation, they agree.

Unfortunately, these four theories disagree about questions of termination. The sequence of approximations even gives different answers when different index sets are used for the sequence. Yet all these theories are sound because they differ about things that are not observable. Furthermore, all sound theories are incomplete concerning questions of termination: for each theory, there are programs whose termination status cannot be decided by the theory. The Halting Problem, which asks for a program (in some programming language) to determine the halting status of all programs (in that same language), is ill-defined by failing to say according to which theory, and is inconsistent by asking for both soundness and completeness.

I have questioned the meaningfulness or worth of specifying termination without a time bound, but I expect many people will cling to the feeling that it is meaningful and worthwhile. I now introduce you to calumation, a word that is not in any dictionary, and is absolutely meaningless. The Calumation Problem is:

> In a general-purpose programming language, write a program that reads two texts (character strings) $p$ and $i$. If $p$ represents a program in that same language whose execution with input $i$ calumates, then output *true*. If $p$ does not represent a program in that same language, or represents a program in that same language whose execution with input $i$ does not calumate, then output *false*.

To "prove" that calumation is incomputable, all we need is one positive example, and one negative example. So let's say that $P$ is a procedure whose execution calumates, and $N$ is a procedure whose execution does not calumate. Assume, for the sake of showing a contradiction, that the calumation function can be programmed, and has been programmed; let's call it *cal*. Then we can write a procedure like this:

**procedure** *caldiag* (*s*: **string**);
**begin**
    **if** *cal* (*s*, *s*) **then** *N* **else** *P*
**end**

Does execution of *caldiag* ('*caldiag*') calumate? If it does, then *cal* ('*caldiag*', '*caldiag*') should return *true*, and so we see from the body of procedure *caldiag* that its execution does not calumate. If it doesn't calumate, then *cal* ('*caldiag*', '*caldiag*') should return *false*, and so we see from the body of procedure *caldiag* that its execution calumates. We have a contradiction (inconsistency). If we accept the standard argument for the incomputability of halting, we must now conclude that the calumation function cannot be programmed; it is incomputable.

We thus "prove" that the meaningless calumation function is incomputable exactly the same way we "prove" that halting is incomputable [3]. In my opinion, a "proof" that proves a completely undefined function to be incomputable is suspicious.

## Conclusion

Termination without a time bound is a worthless property, whether as a guarantee or as a specification (some would even say meaningless). The Halting Problem, which asks for a program (in some programming language) to determine the halting status of all programs (in that same programming language), is ill-defined by failing to say according to which theory the halting status should be determined. But the meaning of "halting" doesn't matter, because the "proof" that halting is incomputable has nothing to do with halting; it works just as well (or badly) "proving" that a meaningless property is incomputable. That is because the "proof" actually shows an inconsistency in the specification that is independent of the property.

**other papers on halting**